\documentclass[12pt]{iopart}
\usepackage{graphicx}
\usepackage{amssymb}
\usepackage{bm}
\begin{document}

\title[Self-consistent renormalization group approach]{Self-consistent
renormalization group approach to continuous phase transitions in alloys.
Application to ordering in $\beta$-brass}

\author
{V I Tokar}

\address{Universit\'e de Strasbourg, CNRS, IPCMS, UMR 7504,
F-67000 Strasbourg, France}
\ead{tokar@ipcms.unistra.fr}
\address{$^1$G. V. Kurdyumov Institute for Metal Physics of the 
N.A.S. of Ukraine, 36 Acad. Vernadsky Boulevard, UA-03142 Kyiv, Ukraine}
\begin{abstract}
	A self-consistent (SC) renormalization group approach of the
	effective medium kind has been developed and applied to the
	solution of the Ising model (IM).  A renormalization group
	equation in the local potential approximation (LPA) derived
	previously for spatially homogeneous systems has been extended
	to the lattice case and supplemented with a self-consistency
	condition on the pair correlation function.  To validate the
	approach it has been applied to the simple cubic IM and good
	agreement of the spontaneous magnetization calculated with the
	use of the SC-LPA equation with the available exact Monte Carlo
	simulations data has been established.	Next the approach has been
	applied to the bcc IM corresponding to $\beta$-brass. With the use
	of the effective pair interaction parameters from available {\em
	ab initio} calculations the critical temperature, the correlation
	length and the long range order parameter in the vicinity of the
	critical point have been calculated in excellent agreement with
	experimental data.  Qualitative and quantitative arguments have
	been given in support of the suggestion that the experimentally
	observed decrease of the effective critical exponent of the order
	parameter in comparison with the universal value is enhanced
	by the positive value of the second neighbour pair interaction
	found in the {\em ab initio} calculations.
\end{abstract} 
\maketitle
\section{Introduction}
Modern theory of alloys aims at describing the order-disorder phase
transitions fully {\em ab initio} without resort to any phenomenological
input \cite{ducastelle}. However, because inclusion of correlated disorder
in the band structure calculations meets with severe difficulties
\cite{elliott_theory_1974,ziman_models_1979}, theoretical treatment
of interatomic correlations at finite temperature is usually done in
two steps.  At the first step the configuration-dependent electronic
structure energy is mapped onto the Ising model (IM) with the effective
cluster interactions (ECIs) between the spins and at the second step
the IM thermodynamics is treated by means of statistical mechanics
\cite{ducastelle,Zunger1994,Zunger2004,turchi-johnson-V1-2,friedel2007,%
friedel2010}.

Currently no universal theoretical techniques efficient at both
stages exist partly because different alloys may exhibit qualitatively
different behaviour and a technique efficient in alloys of one kind
performs poorly in alloys of different kind. In particular, cluster
methods of \cite{ducastelle,tokar_new_1997,tan_topologically_2011}
that proved to be efficient in the description of the first order phase
transitions fail to correctly describe continuous transitions because
small clusters cannot properly account for the long range correlations
in the critical region.  The cluster sizes $N_c$ are restricted to small
values because the number of terms in the equations grows as $2^{N_c}$
so in practical calculations cluster radii have to be bounded by a few
lattice constants \cite{ducastelle,tokar_new_1997,tan_topologically_2011}.

The restriction on the cluster sizes is greatly alleviated in the Monte
Carlo (MC) method \cite{binder} where instead of all spin configurations,
as in the analytical cluster theories, only a relatively small number
of the most important configurations is explicitly simulated. As a
consequence, the linear size $L$ of the simulation box, the homologue of
the clusters in the analytical approaches with $N_c\propto L^3$ in 3D,
is limited only by available computational resources. A major advantage of
the MC approach is that it can treat IM of any complexity. For example, in
\cite{Zunger2004} the Hamiltonian corresponding to Ta-Mo alloy with ECIs
consisting of eight pair and five many-body interactions was simulated
in broad range of temperatures with the use of the simulation boxes with
$L=16-32$ lattice units.  The simulations predicted, in particular,
a yet unobserved continuous ordering transition with the critical
temperature $T_c$ in the range 600--1000~K.  The poor accuracy in $T_c$
determination was due to an unusually broad maximum in the specific heat
curve. Obviously that the critical behaviour of the specific heat was
impossible to describe at this level of accuracy. But the problem was
not only in the complexity of the Hamiltonian or in too small simulation
boxes and insufficient statistics. Even on simple cubic (sc) lattice
and IM with only NN interactions the simulations with box sizes up to
$L=1024$~l.u.\ it proved impossible to determine with good accuracy
the specific heat critical exponent $\alpha$ \cite{ferrenberg2018}. In
earlier study with $L\leq512$ the error in determination of $\alpha$
was $\sim100\%$ \cite{bcc-fcc-diamond-Kc}.

Still, because other critical exponents were
quite accurately determined in the simulations in
\cite{talapov_M(t),bcc-fcc-diamond-Kc,ferrenberg2018}, the MC approach
in principle can be used for the description of experiments on ordering
in $\beta$-brass in \cite{beta-brass2016} because specific heat was
not measured in the study.  However, the MC simulations needed may
require quite extensive computations in order, for example, to determine
differences between the values of the critical exponent of the order
parameter $\beta$ in different experimental set-ups and/or models.
In experiments in \cite{beta-brass2016} the difference in $\beta$
values that we would like to explain was less than $4\%$. But in
\cite{bcc-fcc-diamond-Kc}, where according to the authors the MC data of
unprecedented size were simulated, $\beta$ could be determined with the
accuracy of only $\pm2.5\%$, so the error in the difference between two
values of $\beta$ would exceed the difference $<4\%$ we are interested in.

Much better accuracy was achieved in recent simulations in
\cite{ferrenberg2018} but at the cost of $2\times10^7$ CPU core hours (2.3
thousand years) on five Linux clusters. Rough estimates show that for the
accuracy which would be sufficient for the present study these numbers
could be reduced.  But it should be taken into account that instead
of one nn model on sc lattice studied in \cite{ferrenberg2018} three
different models on bcc lattice with additional next nn interactions in
two of them would need to be simulated. In view of the discussion in the
previous paragraph it should be concluded that large-scale computations
would still be necessary.

The extensive MC simulations seems to be the only practical way to
reliably calculate thermodynamic quantities in the critical region
in the case of Hamiltonians containing many-body ECIs that usually arise
in realistic descriptions of the configurational alloy energy 
\cite{ducastelle,Zunger1994,Zunger2004,turchi-johnson-V1-2,friedel2007,%
friedel2010}.  However, in some alloys the energy can
be described with the use of only pair interactions, in
particular, $\beta$-brass is considered to be such an alloy
\cite{beta-brass2016,turchi-johnson-V1-2}. In this case the
partition function can be represented in the form of a functional
integral over a scalar lattice field with the field Hamiltonian
formally of the Ginzburg-Landau type with conventional non-local
quadratic (``free'') part and a local interaction potential (see,
e.g., \cite{gamma_exp,tokar_new_1997,tan_topologically_2011}). The
critical behaviour of the models of this type can be effectively
treated within the functional renormalization group (RG) approach,
in particular, within the local potential approximation (LPA)
\cite{wilson,local_potential,berges_non-perturbative_2002,%
caillol_non-perturbative_2012}.

The aim of the present paper is to derive a RG equation in the
LPA based on the self-consistent functional formalism developed in
\cite{gamma_exp,tokar_new_1997,tan_topologically_2011}.  Though unlike
the MC method this approach is not universal and systematic, it has
some important advantages. First, it can be formulated directly in
the thermodynamic limit, so no need for repeated simulations with
different $L$ values with subsequent non-trivial interpolation to the
infinite system size needed in MC simulations in the critical region
\cite{talapov_M(t),bcc-fcc-diamond-Kc,ferrenberg2018}.  Second, the pair
interactions of any extent can be treated in exactly the same manner as
nn interactions. Third, the critical behaviour can be described within
well established RG framework so the universality properties and scaling
laws hold for all periodic lattices, unlike in MC simulations where their
validity is not guaranteed \cite{bcc-fcc-diamond-Kc}.  Furthermore,
the cluster MC algorithms that are needed in large-scale simulations
to overcome the critical slowdown degrade their performance in the
presence of competing interactions \cite{blote_cluster_2002}. But
such interactions are ubiquitous in metallic alloys where they
arise due to the Friedel oscillations of the electron density
\cite{friedel2007,friedel2010,Zunger2004}. In the proposed RG approach
such interactions would not pose any complications, as will be seen in the
calculations in the {\em ab initio} IM of $\beta$-brass with competing
nn and next nn interactions \cite{turchi-johnson-V1-2}. Finally, the RG
equation in the LPA is computationally undemanding and can be solved on
practically any computer.

The equation will be obtained by modification of the RG equation in the
LPA derived in \cite{1984} for the Ginzburg-Landau model in homogeneous
space.  The modification will consist in adaptation of the equation to the
lattice case and in imposing the self-consistency condition similar to
that used in \cite{gamma_exp,tokar_new_1997,tan_topologically_2011}.
The equation that will be called the SC-LPA RG equation
belongs to the class of nonperturbative RG equations in the
LPA \cite{wilson,local_potential,berges_non-perturbative_2002,%
caillol_non-perturbative_2012} and shares their known shortages.
In particular, the universal quantities, such as the critical exponents,
are approximately reproduced in the LPA in 3D case but not in 2D;
the non-universal quantities, such as the critical amplitudes, are
accurate to the lowest order in the interaction but in the strong
coupling case that will be of main interest in the present study the
SC-LPA should be considered as a heuristic closed-form approximation.
Formally it is analogous \cite{gamma_exp} to such successful 
approximation as the coherent potential approximation (CPA) and DMFT
\cite{elliott_theory_1974,maier_quantum_2005} which in the strong
coupling case can be justified only in some limiting cases (e.g.,
in infinite dimensions) but have been successfully applied to many
physical problems. Therefore, before proceeding to the description of
$\beta$-brass we will first check and validate the SC-LPA RG equation
by comparing its predictions with known reliable solutions of similar
problems in \cite{talapov_M(t),fisher_theory_1967,liu_fisher89}.
\section{\label{formalism}Formalism}
In the pair approximation the configuration-dependent contribution to
the total energy of an equiatomic binary alloy in the IM formalism reads
\cite{ducastelle,turchi-johnson-V1-2}
\begin{equation}
	E_{conf} = \frac{1}{8}\sum_{ij}V_{ij}s_is_j
	\label{Econf}
\end{equation}
where $s_i=\pm1$ are the Ising spins occupying $N$ lattice
sites $\{i\}$ and $V_{ij}$ are the  effective pair interactions
which following \cite{turchi-johnson-V1-2,beta-brass2016}
we will restrict to only the nearest neighbour ($V_1$)
and the second neighbour ($V_2$) interactions which is
sufficient for the discussion of ordering in $\beta$-brass
\cite{fisher_theory_1967,turchi-johnson-V1-2,beta-brass2016}.
In (\ref{Econf}) linear in $s_i$ terms are absent because the
transformation $s_i\to-s_i$ corresponds to replacement of atoms of one
kind by the atoms of another kind and in the equiatomic alloy this should
not change the configurational energy. In the IM language this means that
the external magnetic field is equal to zero. On bipartite lattices,
such as the simple cubic (sc) and the bcc lattices, this additionally
makes possible to switch the signs of spins $s_i\to-s_i$ on one of the
two interpenetrating sublattices and simultaneously reverse the signs
of $V_{ij}$ connecting spins at different sublattices to arrive at a
model with the same statistical properties but with different order
parameter \cite{fisher_theory_1967}. In the case under consideration
the antiferromagnetic order will change to the ferromagnetic 
one and because the ferromagnetic order is simpler, in the study of ordering in
$\beta$-brass (bcc lattice) we will deal with the transformed system. To
avoid confusion, $V_1$ and $V_2$ will retain their physical values while
in explicit calculations we will use the dimensionless (i.e., divided by $k_BT$)
Hamiltonian of the form
\begin{equation}
	H_I=\frac{1}{2}\sum_{ij}\epsilon_{ij}s_is_j 
	- \sum_ih_is_i+N\epsilon_0/2
	\label{H_I}
\end{equation}
with the interactions between the nn and the second neighbour
spins $\epsilon_1=-{V_1}/{4k_BT}$ (note the sign reversal) and
$\epsilon_2={V_2}/{4k_BT}$, respectively.  Besides, we introduced into
$\epsilon_{ij}$ a diagonal part $-\epsilon_0\delta_{ij}$ with
\begin{equation}
	\epsilon_0=\frac{-8V_1+6V_2}{4k_BT}
	\label{epsolon0}
\end{equation}
which is compensated by the last term in (\ref{H_I})) because of the
identity $s_i^2=1$. This is done to ensure the quadratic behaviour of the
Fourier-transformed $\epsilon$ at small momenta \cite{fisher_theory_1967}
\begin{equation}
	\epsilon({\bf k})|_{{\bf k}\to0}\simeq 
	\frac{V_1-V_2}{k_BT}a^2{\bf k}^2
	\label{k20}
\end{equation}
where $a$ in the bcc case was chosen to be equal to one half of the
length of the cube edge so that the vectors connecting nn sites have
coordinates $(\pm a,\pm a,\pm a)$ and their length $a_1=\sqrt{3}a$
(in the sc case the cube edge and the nn distance coincide).

Besides, in (\ref{H_I}) we added the linear coupling of spins to the
source field $h$ that will be needed, e.g., in the formulation of the
self-consistency condition.  At the end of the calculations, however,
it will be set equal to zero because only the equiatomic alloys and the
IM in zero external field will be studied in the present paper.

The calculations below will be based on the SC approach of the
effective medium type introduced in \cite{gamma_exp}. Because
the formalism was recapitulated in several papers (see, e.g.,
\cite{tokar_new_1997,tan_topologically_2011,tokar2019effective})
only its one-component variant sufficient for IM will be briefly
explained below.  The derivation of the SC condition is based on the
observation that the conventional in the many-body theory separation
of the Hamiltonian into the quadratic (or harmonic) in the fluctuating
field part and an interaction part which is of higher order in the field
is not unique. An arbitrary quadratic term can be added to the harmonic
part and simultaneously subtracted from the interaction part which would
leave the Hamiltonian unchanged. The reason for this transformation
is that the harmonic part defines the propagator of the perturbation
theory and the arbitrary term can in principle be adjusted so that the
propagator was equal to the exact pair correlation function of the field
which arguably is the most useful and most often calculated correlation
function in the many-body theory.

In the functional-integral representation the partition function of the
Ising model (\ref{H_I}) reads
\begin{equation}
	Z[h]=e^{N\epsilon_0/2}\prod_l\int ds_l 2\delta(s_l^2-1)e^{-\frac{1}{2}
	\sum_{ij}\epsilon_{ij}s_is_j+\sum_i h_is_i}
	\label{Z}
\end{equation}
where the delta functions fix the continuous spins $s_i$
to their Ising values $\pm1$. By standard manipulations
\cite{gamma_exp,tokar_new_1997,tan_topologically_2011,tokar2019effective}
(\ref{Z}) can be cast in the form (see, e.g., equations (5) and (6)
in \cite{gamma_exp})
\begin{equation}
	Z[h]=\exp\left(\frac{1}{2}hGh\right)R[Gh]
	\label{Z2}
\end{equation}
where the vector-matrix notation has been used in the $N$-dimensional
space of the lattice sites so that, e.g., $Gh = \sum_jG_{ij}h_j$,
etc. The propagator matrix $G$ is translationally-invariant and its
Fourier transform reads
\begin{equation}
	G({\bf k}) = \frac{1}{\epsilon({\bf k})+r}.
	\label{G}
\end{equation}
Here momentum-independent constant $r$ plays the role of
the SC self-energy in the single-site cluster approximations
\cite{gamma_exp,tokar_new_1997,tan_topologically_2011,tokar2019effective}.
In the IM case it can be introduced into (\ref{Z}) in the same manner
as $\epsilon_0$ in (\ref{H_I}) with the corresponding compensating term
accounted for in (\ref{Z2}) through
\begin{eqnarray}
	R[s]&=&\det(2\pi G)^{1/2}e^{N(r-\epsilon_0)/2}
	\exp\left(\frac{\partial}{\partial s}
	G\frac{\partial}{\partial s}\right) \prod_l[2\delta(s_l^2-1)]\nonumber\\
	&=&\exp\left(\frac{\partial}{\partial s}G\frac{\partial}
	{\partial s}\right)\exp\left(-U^{b}[s]\right).
	\label{R}
\end{eqnarray}
Here we introduced the ``bare'' or initial local potential 
\begin{equation}
	U^{b}[s]=\sum_i u^{b}(s_i) 
	\label{U-u}
\end{equation}
which will be renormalized by the RG procedure and in the LPA is
assumed to remain local throughout the whole course of renormalization
\cite{1984}.  
\subsection{Connection with polynomial models}
From (\ref{R}) it follows that in the case of the IM $u^{b}(x)$
in (\ref{U-u}) formally contains a poorly defined contribution
of the form
\begin{equation}
	u^b(x)=-\ln\delta(x^2-1)+\mbox{(f.i.t.)}
	\label{ubIM}
\end{equation}
where by (f.i.t.) we denoted field- or $x$-independent terms.  This
will not pose problems in the calculation below because, as shown in
\ref{LPAapp}, in the differential RG equation we can use $\exp(-u^{b})$
instead of $u^{b}$ so in explicit calculations only the plain
delta-function will appears. However, in the Wilson theory \cite{wilson}
the local potential $u^b$ in (\ref{U-u}) is usually assumed to be a
polynomial function of its argument.  The IM expression (\ref{ubIM})
is very far from the polynomiality and even the analyticity. This poses
the question on whether the results of the conventional RG approach that
heavily relies on the perturbative expansions requiring the analyticity
of $u^b$ apply to the IM.  The possibility that IM is exceptional from
the RG standpoint has been discussed in the literature and large-scale
MC simulations have been performed in support of this viewpoint (see
\cite{bcc-fcc-diamond-Kc,deng_simultaneous_2003} and references therein).

The ``layer-cake'' renormalization scheme makes possible to
reformulate the above question for the case of lattice models
with local interactions as follows. As has already been discussed
in \cite{tokar2019,tokar2019effective} (see, e.g., section 4.2 in
\cite{tokar2019}), in the lattice case the partial initial renormalization
that has led to our equations (\ref{w}) and (\ref{u_ini}) can be performed
exactly because it amounts to application of the site-diagonal operator
$\exp[(t_0/2)\partial^2/\partial s_i^2]$ to the factors $\exp[-u^b(s_i)]$
in (\ref{R}) individually at each site. The exact expression thus
obtained reads
\begin{equation}
	R[s]=\exp\left(\frac{\partial}{\partial s}\tilde{G}\frac{\partial}
	{\partial s}\right)\exp\left[-\sum_iu(s_i,t_0)\right]
	\label{Rr}
\end{equation}
where $u$ is defined in (\ref{u-lnw}) and (\ref{w}) and
\begin{equation}
	\tilde{G}_{ij}=G_{ij}-t_0\delta_{ij}
	\label{tildeG}
\end{equation}
where the second term on the right hand side is subtracted because it
has already been accounted for in the partial renormalization.

Expression (\ref{Rr}) is exact and is amenable to treatment by means
of the perturbation theory because due to the integration in (\ref{w})
the IM now is represented by $u$ in (\ref{u_ini}) which can be expanded
in a convergent Taylor series. For polynomial theories, e.g., for the
conventional $\phi^4$ the initial function $u$ calculated in (\ref{u-lnw})
and (\ref{w}) will also be representable as the infinite Taylor series
so on the same lattice and the same dispersion $\epsilon$ the difference
between IM and $\phi^4$ theories will be only in the coefficients of the
expansion of the local potential at time $t_0$. Thus, there seems to be
no reasons why the critical properties as described by the RG would be
different in the two cases . Therefore, in the present paper we assume
that the RG theory is fully applicable to IM. This assumption presumes,
in particular, that the MC simulations of IM on the largest lattices
used today cited in \cite{bcc-fcc-diamond-Kc} are not able to correctly
describe the critical behaviour.
\subsection{The self-consistency condition}
As follows from (\ref{Z}), the spin correlation function can be found
by differentiating the logarithm of $Z[h]$ in (\ref{Z2}) with respect to
$h_i$ twice which after setting $h$ to zero gives in the matrix notation 
\cite{gamma_exp,tokar_new_1997,tan_topologically_2011,tokar2019effective}
\begin{equation}
	\left\Vert\langle s_is_j\rangle\right\Vert = 
	G-G\left.\frac{\partial^2 U[s]}{\partial s\partial s}G\right|_{s=0}.
	\label{Gexact}
\end{equation}

Because the exact partition function does not depend on $r$, the value
of the latter can be chosen arbitrarily. In approximate calculations,
however, the independence will usually be lost in which case $r$ can
be used as a free parameter to improve the approximation.  In effective
medium theories, one aims at choosing $r$ in such a way that the second
term in (\ref{Gexact}) disappeared and propagator $G$ coincided with
the exact correlation function. This would mean that the propagation
of an individual (quasi)particle within the medium is unperturbed by
the scattering described by the second term, hence the term ``effective
medium''.

In general, however,  it is impossible to set the second term in
(\ref{Gexact}) to zero with the use of a single parameter because in
this case $r$ would coincide with the exact self-energy $r^{exact}({\bf
k})$ of the system which in non-trivial models depends on ${\bf k}$.
With only one momentum-independent parameter at hand the effective
medium condition can be satisfied only approximately. In the single-site
approximation it is assumed that in the exact renormalized potential
$U^R[s]=-\ln R[s]$ (here and below we will designate by superscript $R$
all fully renormalized quantities) the site-diagonal terms dominate
so similar to (\ref{U-u}) $U^R$ can be approximated by a sum of local
potentials $u^R(x)$ and the condition
\begin{equation}
	u^R_{xx}|_{x=0} =0
	\label{u_xx=0}
\end{equation}
will nullify the second term in (\ref{Gexact}) locally which roughly corresponds
to assuming the site-local self-energy $r\approx r^{exact}_{ii}$.

However, in studying the critical region one is interested mainly in the
behaviour of the long range fluctuations, so with only one free parameter
being available it seems more logical to impose the self-consistency
condition on the self-energy at the smallest value of the momentum
$r\approx r^{exact}({\bf k\to0})$.  As will be shown below, in our SC RG
approach this will amount to imposing condition (\ref{u_xx=0}) on the
local potential $u^R$ obtained as the solution of the LPA RG equation.
\subsection{The SC-LPA RG equation} Finding the partition function
(\ref{Z2}) is equivalent to calculating functional $R$ in (\ref{R}). But
for our purposes it will be sufficient to find only the function of the
homogeneous field $h_i=h=Const$ which in (\ref{Z2}) will be replaced by
$Gh=h/r$ (see (\ref{G})). Of course, before doing this substitution all
partial derivatives in (\ref{R}) should be taken.

Calculation of the derivatives in (\ref{R}) by means of a RG technique in
the LPA has been explained in detail in \cite{1984} (cf.\ our equation
(\ref{R}) with equation (4) in \cite{1984}). A slight difference with
the present case is that in \cite{1984} all Fourier components $s_{\bf
k}$ were set to zero while now we want to preserve the component with
${\bf k=0}$.  This is trivially achieved in equation (8) in \cite{1984}
by simply not setting to zero the argument of the fully renormalized
local potential $u^R(x)$ because after successive elimination of all
higher-momenta components the remaining $x$ corresponds to $s_{\bf k=0}$.
In the present study $x=h/r$ will be retained in order to calculate the
dimensionless free energy per site in the external homogeneous field $h$
needed in the derivation of further thermodynamic quantities
\begin{equation}
	f(h) = -\ln Z = u^R(x)|_{x=h/r} - {h^2}/{2r}.
	\label{f}
\end{equation}
Here the use has been made of equations (\ref{Z2}), (\ref{G}), (\ref{R}) 
and (\ref{U-u}).

More serious problem to resolve is that in \cite{1984} the RG equation
was derived for statistical models in homogeneous isotropic space. Though
it is not difficult to adopt the ``layer cake'' renormalization scheme of
\cite{1984} to lattice models \cite{tokar2019effective}, in the present
paper we adopt another possibility based on the observation made in
\cite{velicky_single-site_1968} in the theory of the single-site CPA.
Namely, in \cite{velicky_single-site_1968} it was shown that for
any single-band density of states (DOS) it is possible to construct
rotationally-invariant dispersion $\tilde{\epsilon}(k=|{\bf k}|)$ that
would reproduce it.  But in \cite{latticeRG2010,tokar2019effective} it
was found that the LPA RG equations depend on the lattice structure
only through the DOS corresponding to dispersion $\epsilon({\bf
k})$.  This means that it should be possible to apply to lattice
systems the LPA equations derived for the isotropic space. The only
problem is that the isotropic dispersion is not uniquely defined
\cite{velicky_single-site_1968}. However, as we show in \ref{LPAapp},
$\tilde{\epsilon}$ can be completely excluded from the LPA equation of
\cite{1984} by a change of the evolution parameter from the momentum
cut-off $\Lambda$ to ``time'' $t$ defined in (\ref{t}) thus avoiding
the ambiguity. Specifically, by substituting (\ref{dt}) and (\ref{p})
in (\ref{Eq8}) one gets
\begin{equation}
	u_t = \frac{1}{2}\left[p(t)u_{xx}
	- u_x^2\right].
	\label{LPA}
\end{equation}
where the subscripts denote the partial derivatives and 
\begin{equation}
	p(t)=D_{tot}(t^{-1}-r)=\int_0^{t^{-1}-r}dE D(E)
	\label{p-latt}
\end{equation}
in complete agreement with $n=1$ lattice case in
\cite{tokar2019effective}.  We note that in the ferromagnetic case under
consideration the integration over $t$ in (\ref{LPA}) is bounded from
above because $D(E)$ in (\ref{p-latt}) vanishes at negative $E$ where
dispersion $\epsilon({\bf k})$ is equal to zero so when $t$ exceeds
$r^{-1}$, $p(t)$ also turns to zero. Thus, unlike in more conventional
LPA approaches \cite{local_potential,caillol_non-perturbative_2012}
in \cite{1984} and in the SC LPA the evolution spans a finite interval
of $t$ values except at the critical point where $t_{max}=r^{-1}$
becomes infinite.

The RG equation in the LPA (\ref{LPA}) with the initial condition
(\ref{u_ini}) and the self-consistency condition (\ref{u_xx=0}) constitute
the SC-LPA RG scheme that will be used in explicit calculations throughout
the present paper.

Equation (\ref{LPA}) could be readily integrated numerically in the
symmetric phase above $T_c$.  Below $T_c$, however, insurmountable
numerical difficulties have been encountered. The problem has been
attributed to the exact quadratic partial solution (the Gaussian model)
which in the coexistence region below $T_c$ becomes negative and singular
at some point $t_{sing}>0$:
\begin{equation} 
	u^{G}({ x},t)=\frac{{ x}^2}{2(t-t_{sing})}+\mbox{(f.i.t.)}
	\label{u_G}
\end{equation} 
At $t=0$ the initial curvature in solution (\ref{u_G}) is negative
and diverges to $-\infty$ as $t\to t_{sing}$.  The integration cannot
go beyond this point because the singularity is non-integrable, so it
was identified with the end point of the integration $t_{sing}=1/r$.
The singularity, however, is not a deficiency of the LPA.  In fact,
it ought to be expected on physical grounds because the magnetic
susceptibility should be infinite in the coexistence region but according
to (\ref{Z}) and (\ref{f}) it is given by the second derivative
\begin{equation}
	\chi = {d^2\ln Z}/{dh^2}|_{h=0} = 1/r-u^R_{xx}|_{h=0}/r^2.
	\label{chi-u}
\end{equation}
And because $r$ in (\ref{chi-u}) is proportional to the squared inverse
correlation length $r\propto\xi^{-2}$ it should be finite both above
and below $T_c$. Hence, the infinite susceptibility can arise only
from the second term so its unboundedness is dictated by the physics of
the problem.

The physical soundness of the approximation is gratifying but we have
to find a way of dealing with the singularity. In view of the direct
connection between $u$ and the free energy (\ref{f}), a plausible idea
would be to resort to a Legendre transform (LT) of $u(x)$, say, $v(y)$,
because under the transform the second derivatives of $v$ and $u$ would
be inversely proportional to each other \cite{zia} and the infinity
in $u_{xx}$ would turn into numerically manageable zero in $v_{yy}$.
This general idea has been realized in a non-canonical way via a
LT-like $t$-dependent transform explained in \ref{legendre-like} which
for simplicity we will continue to call the LT transform.  Equations
(\ref{v-u}) and (\ref{y-x}) have been obtained as a generalization
of the $t$-independent LT suggested in \cite{x-yLegendre} (see also
\cite{local_potential}). Though our LT does not have the canonical form
\cite{zia}, it solves the singularity problem because the transformed
LPA equation
\begin{equation}
	v_t = \frac{p(t)v_{yy}}{2(1+\bar{t} v_{yy})}  \label{LPA2}
\end{equation}
($\bar{t}=t-t_0$) obtained from (\ref{vt-ut}) and (\ref{uxxvyy}) has a
Gaussian solution with a $t$-independent quadratic in $y$ term which thus
is non-singular in $t$.  In the coexistence region where $t_{sing}=1/r$
the LT-transformed (\ref{u_G}) reads
\begin{equation}
	v_G(y,t) = -\frac{y^2}{2\bar{t}^R }+\mbox{(f.i.t.)}  \label{v_G}
\end{equation}
where $\bar{t}^R=1/r-t_0$. With the use of (\ref{uxxvyy}) susceptibility
(\ref{chi-u}) expressed in the $v-y$ variables is
\begin{equation}
	\chi=\frac{1}{r}-\left.\frac{v_{yy}^R}{r^2(1+\bar{t}^R v_{yy}^R)}
	\right|_{h=0}.
	\label{chi-v}
\end{equation}
As is seen, though the solution (\ref{v_G}) in the coexistence region
is finite, the susceptibility in (\ref{chi-v}) is infinite, as needed.

It is to be noted that (\ref{u_G}) and (\ref{v_G}) are only particular
solutions of the RG equations and there is no obvious reason why
they should dominate the solution for arbitrary non-Gaussian models,
especially taking into account that in the disordered phase the solution
$u$ or $v$ for the IM are non-Gaussian. Nevertheless, in the numerical
solutions of the IM in the coexistence region the $y$-dependent part
of $v^R$ was indistinguishable from $v^G$ within the accuracy of the
calculations which was $O(10^{-6})$ in our case (see \ref{numerics}).
This is illustrated in figure \ref{fig7}.

A minor inconvenience of dealing with the LT variables $v$ and $y$
is that they do not have an obvious physical meaning. But in view of
(\ref{vy-ux}) the transform can be easily inverted, so that at the end
of the integration we have
\begin{eqnarray}
  \label{h}
  && x = h/r = y+\bar{t}^R  v^R_y \\ 
  \label{u(y)}
  && u^R = v^R+\bar{t}^R  v_y^2/2.
\end{eqnarray}
Thus, in view of (\ref{f}), both the field and the free energy can
be represented in parametric form in terms of $y$ and $v$ so other
thermodynamic quantities can be expressed through $y$ and $v$ with the
use of the standard thermodynamic relations.

The discussion of the LPA solution below $T_c$ will be continued in
section \ref{ordered} but first let us consider the simpler disordered
phase.
\section{\label{disordered}Disordered phase}
Using the IM language, in experiments above $T_c$ in \cite{beta-brass2016}
the authors measured the spin-spin correlation function and compared
it with that calculated in \cite{fisher_theory_1967} for the nn bcc IM.
In slightly modified notation ($a_1$ instead of $a$), expression (11.1)
for the susceptibility in \cite{fisher_theory_1967} reads
\begin{equation}
	\hat{\chi}({\bf k},T)=\frac{({a_1}/{r_1})^{2-\eta}}
	{[(\kappa_1a_1)^2+a_1^2K^2({\bf k})/(1-\eta/2)]^{1-\eta/2}}.
	\label{chi_FB}
\end{equation}
Definitions of quantities entering this expression can be found in
\cite{fisher_theory_1967} so here we only note that $\kappa_1$ is the
inverse correlation length $\xi^{-1}$ and that $r_1$ in (\ref{chi_FB})
is unrelated neither to our $r$ nor to the nn sites.

In the LPA the susceptibility is given by $G({\bf k})$ (\ref{G}) which,
in particular, means that $\eta=0$ \cite{local_potential,1984} and in
this approximation it should coincide with (\ref{chi_FB}).  To cast
the two expressions in the same form we first multiply the numerator
and the denominator of (\ref{G}) by $a_1^2$ and note that $K^2({\bf
k})$ in (\ref{chi_FB}) was defined in \cite{fisher_theory_1967} as our
$\epsilon({\bf k})$ but normalized so that at small $k$ it behaved as
$k^2$ with the coefficient unity. In view of (\ref{k20}) this means that
(\ref{G}) will acquire the form of (\ref{chi_FB}) if we further divide
the numerator and denominator by $(V_1-V_2)a^2/k_BT$. By comparing the
numerators one finds that in the LPA:
\begin{equation}
	\left(\frac{a_1}{r_1}\right)^2\simeq\frac{3k_BT}{V_1-V_2}
	=\frac{3T^{r}}{4(1-V_2/V_1)} \label{r1}
\end{equation} 
where use has been made of the fact that $(a_1/a)^2=3$ and
for simplicity we introduced the dimensionless temperature
\begin{equation}
	T^{r}=\epsilon_1^{-1}=(V_1/4k_BT)^{-1}. 
	\label{Tdl}
\end{equation}
Similar comparison of the first terms in the denominators gives
\begin{equation}
	(\kappa_1a_1)^2 =(a_1/\xi)^2 \simeq 3T^dr/4(1-V_1/V_2).\label{kappa1}
\end{equation} 
The values of parameters in (\ref{chi_FB}) were given by expressions
(9.9) and (9.14) in \cite{fisher_theory_1967}:
\begin{equation}
	\log_{10}\kappa_1a_1\simeq\nu\log_{10}\tau+B_0+B_1\tau \label{BB}
\end{equation} 
and
\begin{equation}
	r_1(T)/a_1=(r_1/a_1)_c(1-c\tau+\cdots) \label{r1c},
\end{equation} 
respectively, where $\tau=|1-T/T_c|$ and other parameters are listed in tables 
VI and VII in \cite{fisher_theory_1967}.

\begin{figure}
	\centering \includegraphics[scale=0.8]{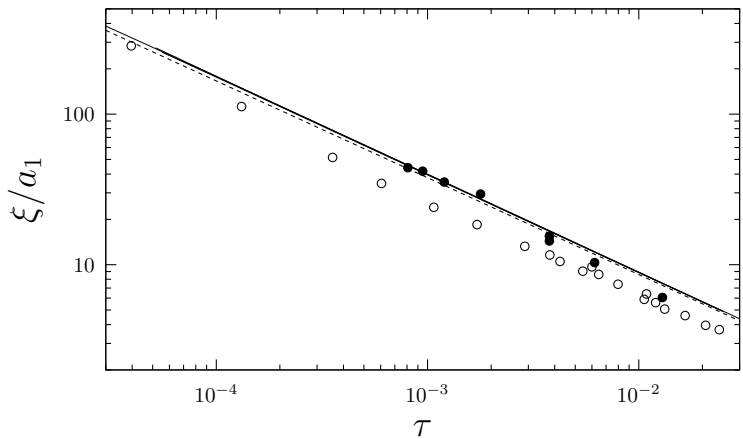}
	\caption{\label{fig1} Correlation length above $T_c$:
	dashed curve and the empty circles were calculated as
	$(\kappa_1a_1)^{-1}$ on the basis of (\ref{BB}) with the
	parameters from \cite{fisher_theory_1967} and from the
	experimental data of Fig.\ 10 in \cite{beta-brass2016},
	respectively; three overlapping solid lines are the LPA solutions
	for the nn IM and for the IM with $V_1=3.8$~mRy and two second
	neighbour interactions $V_2=0.9$~mRy \cite{turchi-johnson-V1-2}
	and -1~mRy; the black circles are experimental data from Fig.\
	9 in \cite{beta-brass2016}.}
\end{figure}

For quantitative comparison, the SC-LPA equation (\ref{LPA2}) was solved 
numerically for nn bcc IM in the vicinity of $T_c$ in the disordered phase. 
Details of the numerical techniques used are given in \ref{LPAapp}. 
The results are compared with the solution of \cite{fisher_theory_1967}
in figure \ref{fig1} and in table \ref{comparison}. 
\begin{table} \caption{\label{comparison}Comparison
of the parameters entering expressions (\ref{BB}) and (\ref{r1c})
for nn Ising model on bcc lattice as calculated in the present work
and in \cite{fisher_theory_1967} (tables VI and VII). For simplicity,
all numbers were rounded so as that the discrepancy between the two
approaches were in one significant figure.} 
\begin{tabular}{@{}lccccccc} \hline
&$\nu$&$B_0$&$B_1$&$\eta$&$(r_1/a_1)_c$&$c$&$(a_1/r_1)_c^{2-\eta}$\\
\hline
		       LPA&0.65&0.352&-0.3&0&0.46&0.50&4.75\\
\cite{fisher_theory_1967} &0.64&0.351&-0.1&0.06&0.45&0.47&4.77\\ \hline
\end{tabular} 
\end{table} 
As is seen, the largest discrepancy is between the values of $B_1$. But
because in (\ref{BB}) $B_1$ is multiplied by $\tau$, at the largest
value of $\tau$ in Fig.\ \ref{fig1} it introduces the error amounting
to only about 1\% of $B_0$. As a result, in this range the discrepancy
between the LPA values of $\xi$ and the values calculated on the basis of
(\ref{BB}) with the parameters from Table VI in \cite{fisher_theory_1967}
is smaller than 3\%.  When $\tau\to0$ the error becomes negligible so the
discrepancy seen in the figure at small $\tau$ should be attributed to the
difference in $\nu$, as can be seen from a steeper LPA curve. Thus, with
the overall discrepancy in a few percent the agreement can be deemed to
be satisfactory taking into account that the authors assess the accuracy
of (\ref{chi_FB}) in 12\% \cite{fisher_theory_1967}. Besides, the LPA
$T_c^{r}$ agreed with the best known estimates for $T_c$ in the nn bcc
IM within 0.3\% \cite{tokar2019effective}.

It is reasonable to assume that the accuracy of the LPA similar
to the nn case will also hold for IM with not too large second
neighbour interactions. This is further confirmed by the fact that
$T_c\simeq748.5$~K calculated with the {\em ab initio} values of
the interactions taken from Fig. 1 in \cite{turchi-johnson-V1-2}
$V_1=3.8$~mRy and $V_2=0.9$~mRy differed from the experimental value
739~K on 1.3\% which is appreciably larger than 0.3\%. This discrepancy
can be a consequence of the further neighbour interactions neglected
in \cite{turchi-johnson-V1-2}.  The correlation length calculated
with these parameters is also shown in figure \ref{fig1} and is almost
indistinguishable from the nn case. This means that according to our
calculations the model can describe the experimental data in disordered
phase as well as the nn model but, in addition it can predict $T_c$
with reasonable accuracy and, besides, has a firm {\em ab initio}
foundation \cite{turchi-johnson-V1-2}.

To complete the check on the influence of the next neighbour interactions
on the behaviour of the correlation length, a model with negative $V_2$
has been solved and also did not show appreciable deviations from the
nn case.  Thus, our calculations do not support the suggestion made in
\cite{dietrich_temperature_1967} that farther-neighbour interactions can
be responsible for disagreement of the nn model with experimental data.
\section{\label{ordered}Ordered phase} 
The behaviour of the order parameter below $T_c$ measured in $\beta$-brass
in \cite{beta-brass2016} is more difficult to interpret quantitatively.
Theoretically, in a close vicinity of the critical point the order
parameter follows the power law
\begin{equation}
	m_0^*(\tau)=a_0^*\tau^{\beta^*},
	\label{power_law}
\end{equation}
where the order parameter $m_0$, the amplitude $a_0$ and the critical
exponent $\beta$ have been starred because in the finite temperature
range they do not correspond to the true critical quantities but only to
the effective ones that are influenced by the corrections to scaling and
cannot be defined independently of the experimental set-up in which
they were measured.

In the Ising universality class the order-disorder transitions
are described by the universal critical order parameter exponent
which value according to the most advanced MC simulations is
\cite{talapov_M(t),bcc-fcc-diamond-Kc,ron2017,ferrenberg2018}
\begin{equation}
	\beta \simeq 0.325\pm0.002.
	\label{beta}
\end{equation}
In this paper we will neglect distinctions between the true
universal value, the value $\beta$ found in \cite{talapov_M(t)}
and $\beta^{LPA}=0.325$ because their differences are negligible on
the scale of variation of $\beta^*$ in our calculations and in
\cite{beta-brass2016} where (\ref{power_law}) fitted to experimental
data in two temperature intervals gave the following values
\begin{eqnarray}
	\label{small_tau}
&&\beta^*=0.313,\quad\quad\quad\quad \tau\lesssim0.014\\
	\label{large_tau}
&&\beta^*=0.29\pm0.01,\;\;\quad \tau\lesssim0.04 
\end{eqnarray}
which disagree with (\ref{beta}) in the second and even in
the first significant digit. Moreover, (\ref{small_tau}) and
(\ref{large_tau}) violate the scaling relation $\beta=\nu(1+\eta)/2$
holding in 3D \cite{RG2002review} because with $\nu=0.64$ adopted in
\cite{beta-brass2016} the universal value of $\beta$ should exceed $0.32$
for any $\eta>0$. 

The above discrepancies should be expected because the power laws
with the universal exponents are strictly valid only asymptotically when
$\tau\to0$ and cannot describe data on finite temperature intervals where
the true behaviour is different from (\ref{power_law}) and in 3D case is
unknown. Rigorous RG theory predicts an infinite number of correction
terms of the power-law type with known exponents but not amplitudes
\cite{wegner_corrections_1972}.  Thus, expression (\ref{power_law}) is
not valid on any finite temperature interval and the quantities entering
it do not have much physical meaning so they were marked by the stars
to distinguish them from the physical spontaneous magnetisation $m_0$,
the critical amplitude $a_0$ and the universal critical exponent $\beta$
that will be calculated below with the use of the SC-LPA RG equation.

Nevertheless, because in \cite{beta-brass2016} some experimental data were
fitted to (\ref{power_law}), below we discuss peculiarities of such a
fit with the reservation that fitting to an incorrect expression is a
poorly defined problem and the fit results will depend on practically all
details of the fitting procedure, such as the distribution of the measured
points, their weights, etc.  The main goal pursued by the nonperturbative
RG approach is to calculate all quantities of interest directly without
the need to resort to heuristic expressions of unknown validity.

As explained in \ref{numerics}, below $T_c$ two quantities should be
determined self-consistently within the SC LPA approach: self-energy
$r$ and the spontaneous magnetisation $m_0$.  With known $v^R(y)$ the
latter according to (\ref{Z}), (\ref{f}), (\ref{vy-ux}) and (\ref{h})
can be calculated as
\begin{equation}
	m_0 = y_0-t_0v^R_y|_{y_0}=y_0/(1-rt_0)
	\label{m0}
\end{equation}
where the last equality was obtained from (\ref{h}) for $h=0$.  But for
general $h\not=0$ the first equation in (\ref{m0}) should be used.

Because the phases above and below $T_c$ are physically quite different,
in the absence of quantitative criteria of the accuracy of the approach
the SC-LPA solution should also be tested in the ordered phase by
comparing it with reliable reference data. To this purpose the highly
accurate MC simulations in the ordered phase made in \cite{talapov_M(t)}
have been used. Though the model studied was the nn sc IM, similar
to bcc the sc lattice is bipartite and its coordination is only 25\%
smaller so the accuracy of LPA in this system should be similar to what
can be expected in the bcc case.
\subsection{\label{sc}Ordering on the sc lattice}
The results of the MC simulations in \cite{talapov_M(t)} were 
summarized in the form of an interpolation formula
\begin{equation}
m_0(\tau) = \tau^\beta(a_0-a_1\tau^{\theta}-a_2\tau)
	\label{talapov}
\end{equation} 
where the precise parameter values are given in \cite{talapov_M(t)};
the rounded values are given in table \ref{a0-a2} below.  In figure
\ref{fig2} magnetisation (\ref{talapov}) is compared with the LPA
calculations.  The accuracy of expression (\ref{talapov}) is quite high,
of order of $10^{-5}$ \cite{talapov_M(t)} but the accuracy of the LPA
calculations were at best $2\cdot10^{-3}$ at $m_0=1$ because of the finite
differentiation step used.  Therefore, in all our fits and figures the
smallest $m_0$ was chosen to be 0.2 in order to have the accuracy at least
not worse than 1\%, though the LPA equation could be easily solved for
much smaller magnetisations.  Good accuracy of the data, however, is vital
for our purposes because we intend to study quantitatively the deviations
of the LPA data from the linearity in the region $\tau\leq0.04$ where they
are hardly discernible on the scale of the graph in figure \ref{fig2}.
\begin{figure}
	\centering \includegraphics[scale=0.8]{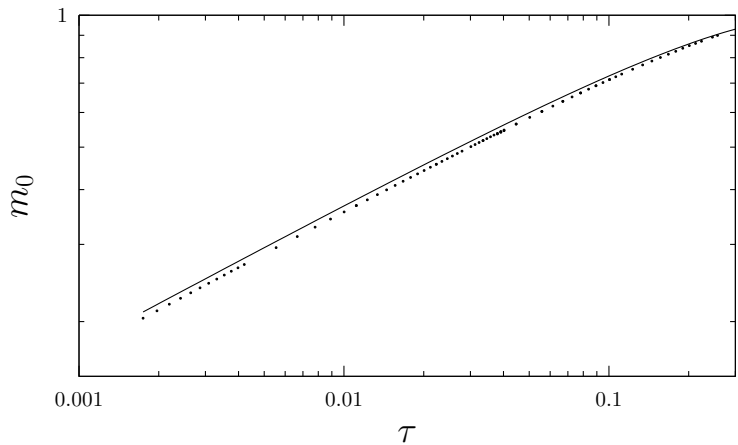}
	\caption{\label{fig2} Spontaneous magnetisation in the sc
	IM: solid line---the interpolation of the exact MC simulations
	(\ref{talapov}) \cite{talapov_M(t)}, symbols---the LPA solution.}
\end{figure} 
The non-linearity of the logarithm of $m_0(\tau)$ is, however, obvious
from the fitting expression (\ref{talapov}).  To assess the quality of
the LPA solution it was fitted to the LPA points in figure \ref{fig2}
with the use of the LPA order parameter exponent $\beta=0.325$
and the leading correction exponent taken to be $\theta=0.5$
\cite{wegner_corrections_1972} because the corrections to it are of
higher order of the $\varepsilon$-expansion \cite{wilson} than
the LPA which is accurate only to the first order in $\varepsilon$
\cite{local_potential,1984}.  As can be seen from table \ref{a0-a2},
similar to the disordered case the worst agreement is with a correction
term, this time with $a_1$ which is about one third smaller than the MC
value. Still, the largest error in $m_0$ introduced by this discrepancy
is about 3.6\% at the maximum value of $\tau=0.26$. It is even smaller
at 0.04 and shrinks to zero as $\tau\to0$. This, however, is an important
difference to us because of the strong influence of the leading correction
on the effective order parameter exponent $\beta^*$ \cite{RG2002review}.
\begin{table}
	\caption{\label{a0-a2} Parameters of the LPA $m_0(\tau)$ fit to
	(\ref{talapov}) with $\beta$ and $\theta$ held fixed compared
	to rounded values from \cite{talapov_M(t)}.}
\begin{tabular}{@{}lcccccc} \hline
    &$T_c$ &$\beta$ &$\theta$ &$a_0$ &$a_1$ &$a_2$\\
\hline 
LPA &4.475 &0.325   &0.50      &1.62  &0.22  &0.41\\
\cite{talapov_M(t)} &4.512 &0.327 &0.51 &1.69&0.34&0.43\\ \hline
\end{tabular} \end{table} 

In \cite{beta-brass2016}, however, the data were fitted not to
(\ref{talapov}) but to more conventional power law (\ref{power_law}) so
let us find out how accurately the SC-LPA reproduces such fits.  As was
already pointed out, in the fit to an incorrect function all details
of the fitting procedure may influence the results. Therefore, because
in \cite{beta-brass2016} the authors fitted a quantity proportional to
$m_0^2$, in checking the reliability of SC-LPA we fitted the squared
power law (\ref{power_law}) to the squared MC data (\ref{talapov})
by minimizing the integral
\begin{equation}
	I = \int_{\tau_0}^\tau [m_0^*(\tau^\prime)^2-m_0(\tau^\prime)^2]^2
	d\tau^\prime
	\label{I}
\end{equation}
with respect to $a_0^*$ and $\beta^*$. In (\ref{I}) it is implicitly
assumed that all data have the same weight and, besides, are homogeneously
distributed within the interval $[\tau_0,\tau]$. These assumptions, of
course, are rather arbitrary but they will allow us to roughly estimate
the span of variation of possible values of $\beta^*$.

\begin{figure}
	\centering \includegraphics[scale=0.8]{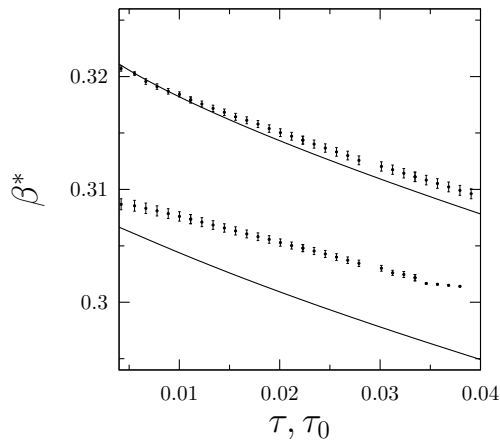}
	\caption{\label{fig3} The effective order parameter
exponent $\beta^*$ fitted to the MC (solid lines) and to the LPA (symbols)
simulation data. For details of the fitting procedure see the text.}
\end{figure}

The integrals in (\ref{I}) can be calculated analytically and the
parameters found exactly. The fitted values of $\beta^*$ are shown in
figure \ref{fig3} by the solid lines. The upper line corresponds to the
fit when the lower limit of integration in (\ref{I}) was held fixed at
$\tau_0^{min}$ corresponding to $m_0=0.2$ while the upper limit varied
from $\tau_0$ to $\tau^{max}=0.04$.  At the lower line the upper limit
was fixed while $\tau_0$ varied from $\tau_0^{min}$ to $\tau^{max}$. This
case roughly imitates the situation when the data at small $m_0$ are
given very low weight because of larger errors.

Similar procedure was applied to the LPA data except that instead of
the integral the sum over discrete points was used in the expression for
$I$. As can be seen in figure \ref{fig3}, the agreement with the fit to MC
data is not perfect and the LPA values show smaller deviations from the
universal $\beta$. This reflects the smaller amplitude of the leading
correction $a_1$ noted above. But it should be born in mind that the
deviation of $\beta^*$ from $\beta$ that we are interested in is less than
10\% and in the worst case of agreement the LPA still predicts $\sim80\%$
of it. The important conclusion from these fits is that the deviations are
similar in magnitude to those obtained experimentally and so potentially
may explain them if the bcc case exhibits deviations of similar magnitude.
\subsection{Ordering on bcc lattice.}
\begin{figure}
	\centering \includegraphics[scale=0.8]{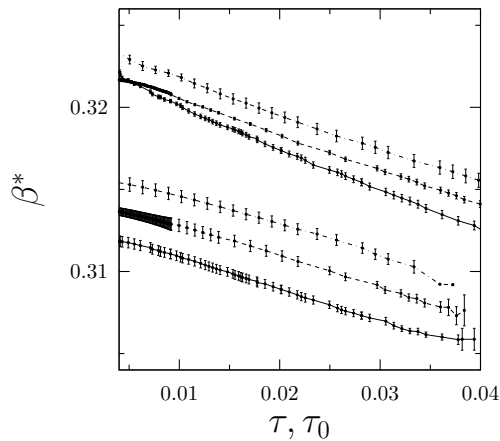}
	\caption{\label{fig4} The upper and the lower groups
	of curves were obtained in as in figure \ref{fig3}. The
	simulated models were the same as listed in the caption to figure
	\ref{fig1}; the solid line corresponds to $V_2=0.9$~mRy,
	dashed line to $V_2=0$ and dashed-dotted line to $V_2=-1$~mRy.}
\end{figure}

Thus, judging from the sc IM, the accuracy $\sim20\%$ may be expected
in the LPA deviations of the effective exponent $\beta^*$ in the bcc
case shown in figure \ref{fig4}.  The simulations were carried
out for the same models as in section \ref{disordered} but this time
the difference between the three cases was clearly visible, though it
was not large.

The important observation that can be made from figure \ref{fig4}
is that the fits seem to support the {\em ab initio} model of
\cite{turchi-johnson-V1-2} in comparison with the nn IM ($V_2=0$) used
in the interpretation of experimental results in \cite{beta-brass2016}.
The difference between the two cases, however, is rather small, not
exceeding the LPA errors estimated in the sc case so the question arises
of whether the difference is real. Because the observation is one of
the main results of the present study, below are given qualitative
arguments in favour of the conclusion that $\beta^*(V_2>0)$ should indeed
be smaller than $\beta^*(V_2=0)$.

To begin with, let us consider a ferromagnetic IM with interactions of the form
\begin{equation}
	\epsilon_{i\not=j}= C(\lambda)e^{-|i-j|/\lambda}
	\label{lambda}
\end{equation}
where $|i-j|$ is the Euclidean distance between the sites, $\lambda$
is a characteristic interaction range and $C<0$ can be chosen so
as to keep the critical temperature fixed, though the latter is
not obligatory.  It is important to note that the nn IM belongs to
the class of models (\ref{lambda}) with $\lambda\to0$. As is known,
in the limit $\lambda\to\infty$ model (\ref{lambda}) tends to the
exactly solvable mean-field (MF) model with all $\epsilon_{i\not=j}$
being equal and the critical exponent $\beta_{MF}=0.5$.  The latter,
however, holds only when $\lambda=\infty$. At any finite $\lambda$ the
model belongs to the same Ising universality class as the nn IM but as
$\lambda$ grows the true critical region shrinks and outside of it the MF
behaviour dominates. Thus, when fitted to the power law (\ref{power_law})
within a finite temperature interval the effective $\beta^*$ should grow
from its initial value close to 0.3 corresponding to nn IM (see figure
\ref{fig4}) toward the MF value 0.5.

Model (\ref{lambda}) is of interest to us because our model with
$V_2=-1$~mRy can be accurately represented by (\ref{lambda}).  Indeed,
with $|V_2/V_1|\approx0.26$ and the distances between the second
and the first neighbours differing on $\simeq0.27a$, the effective
interaction range can be found to be $\lambda\simeq0.2a$. The third
neighbour interaction in this case according to (\ref{lambda}) has the
strength $\sim1.6\%$ of $V_2$ so to a good approximation $V_{l\ge3}$
can be neglected. Obviously, the models with negative $V_2$ but with
smaller $|V_2|$ can be approximated by (\ref{lambda}) even better. Now,
because in the short-range ferromagnetic models belonging to the Ising
universality class, there is no other critical or otherwise singular
points, it should be expected that the behaviour of $\beta^*(\lambda)$
would be monotonous with larger $\lambda$ meaning larger $\beta^*$.

Thus we have shown that for a finite temperature interval near $T_c$
and the models with only nn and the second neighbour interactions the
effective order parameter exponent fitted at this interval should
monotonously diminish from the value $\beta^*(V_2=-1$~mRy) toward
$\beta^*(V_2=0)$. Now by continuity arguments it can be concluded that
when $V_2$ grows farther by acquiring positive values the decrease
of $\beta^*$ should persist which qualitatively agrees with the fits
shown in figure \ref{fig4}. Of course, if $V_2$ becomes sufficiently
large to cause the frustration effects the continuity may fail. But
$V_2=0.9$~mRy is rather small in comparison with $V_1=3.8$~mRy so the
continuity arguments should hold.
\subsection{Ordering in $\beta$-brass.}
\begin{figure}
	\centering
	\includegraphics[scale=0.8]{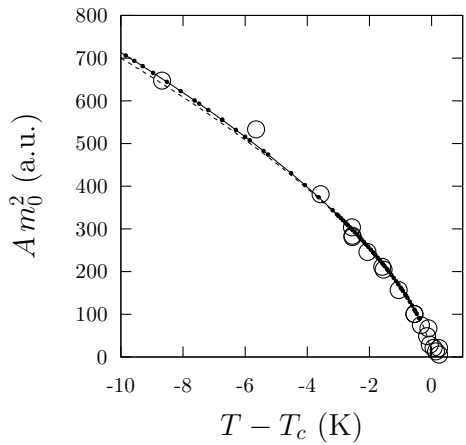}
	\caption{\label{fig5} The empty circles and the dashed curve
are, respectively, the experimental data and their power-law fit
(\ref{power_law}) with $\beta^*=0.313$ taken from Fig.\ 6 in
\cite{beta-brass2016}. The black dots (connected by solid line 
for better visibility) are the LPA results adjusted to the data via 
parameter $A$.}
\end{figure}

The values of fitted $\beta^*$ shown in figure \ref{fig4} indicate
that the experimentally observed behaviour of the order parameter for
$\tau\lesssim0.014$ interpolated in \cite{beta-brass2016} by the power law
(\ref{power_law}) with $\beta^*=0.313$ should be amenable to description
by the SC-LPA equation with the {\em ab initio} parameters $V_1$ and $V_2$
as in \cite{turchi-johnson-V1-2}. Indeed, as shown in figure \ref{fig5},
in this region the LPA calculations compare well with the experimental
points and the power law curve from \cite{beta-brass2016}.
This, however, does not mean that both descriptions are equally adequate.
In contrast to the phenomenological theory of \cite{beta-brass2016},
in the RG approach the problems with the universality and the
scaling relations do not arise \cite{wilson,RG2002review} and
the LPA preserves these features, though with approximate values
of critical exponents ($\beta^{LPA}=0.325$ and $\nu^{LPA}=0.65$)
\cite{local_potential,1984,caillol_non-perturbative_2012}.The small
deviations from the best known values are expected to be corrected
in the future with the use of techniques developed in the theory of
nonperturbative RG \cite{berges_non-perturbative_2002}. Our use of
the rotationally-invariant formalism to describe lattice models should
considerably facilitate the task.  But the main advantage of the SC-LPA
is that there is no need in heuristic expressions to fit experimental
data because all observable quantities can be calculated directly.

Farther from $T_c$, however, in the interval $\tau\lesssim0.04$
the experimentally found value $\beta^*=0.29\pm0.01$ can hardly
be reproduced in the LPA because the effective beta range in figure
\ref{fig4} extends from $\sim0.305$ upwards. The LPA values of $m_0^2$
calculated at the seven $\tau$ points close to those in the inset in Fig.\
11 in \cite{beta-brass2016} fitted to the power law (\ref{power_law})
have given $\beta^*\approx0.315\pm0.002$ which is noticeably greater
than (\ref{large_tau}). Because LPA overestimates $\beta^*$, the real
discrepancy may be smaller but if the sc case is representative of what
may be expected on the bcc lattice, the downward shift of $\beta^*$
in figure \ref{fig3}) would be too small to explain the remaining
discrepancy $\Delta\beta =0.025$.

Thus, the value of effective $\beta^*=0.29$ cannot be quantitatively
understood within the model with parameters of \cite{turchi-johnson-V1-2}.
Two possible explanations for this failure can be envisaged. First, at
$\tau=0.04$ the calculated long range order parameter reaches as large
value as 0.53 which may influence the interatomic interactions propagated
via the electronic subsystem and thus change the values of $V_1$ and $V_2$
as well as introduce additional effective cluster interactions.

\begin{figure}
	\centering
\includegraphics[scale=0.8]{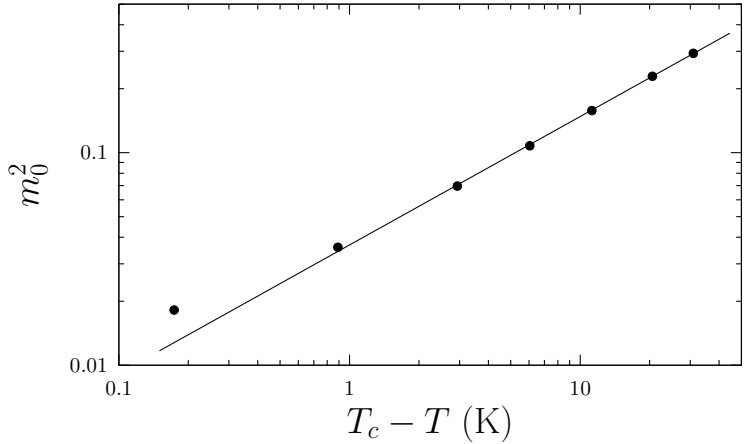}
\caption{\label{fig6} LPA data points (symbols) shifted on
$\simeq0.2$~K toward $T_c$ fitted to power law (\ref{power_law}) (solid
line) with $\beta^*\simeq0.30$; experimental points at these temperatures
were fitted in \cite{beta-brass2016} with $\beta^*\simeq0.29\pm0.01$.}
\end{figure}

The explanation may also lie in the experimental uncertainties in the
temperature measurement in \cite{beta-brass2016} which according to the
authors were of the order of 0.2~K. To assess possible implications,
let us assume that away from $T_c$ the measured temperatures were
systematically overestimated so that they were effectively shifted
toward the critical temperature being about $\Delta T\approx0.2$~K
closer to $T_c$ than they were in reality.  Alternatively, this may
be a consequence of the error in determination of $T_c$, or errors
of both kinds could contribute to the shift.  Now by fitting the same
seven LPA points as above to the re-defined $\tau=1-(T+\Delta T)/T_c$ in
(\ref{power_law}) one finds $\beta^*=0.302\pm0.005$ which already overlaps
with (\ref{large_tau}).  Taking into account that LPA overestimates the
effective exponents the agreement with experiment may be even better.
As can be seen in figure \ref{fig6}, qualitatively the fit looks as good
as the corresponding fit in \cite{beta-brass2016} except at the point
closest to $T_c$. But this point is one of the many in the vicinity
of $T_c$ which are rather scattered due to the steepness of the order
parameter in this region and the perfect agreement of the point with
the fitting curve could be accidental.
\section{Conclusion}
In this paper a SC RG equation in the LPA has been derived
and applied to an accurate quantitative description of the
MC simulation data on the magnetic ordering in the sc lattice
\cite{talapov_M(t)} and to experimental data on the ordering in
$\beta$-brass \cite{beta-brass2016}. In the latter case it has been
shown that with the use of the {\em ab initio} values of the effective
pair interactions \cite{turchi-johnson-V1-2} it has been possible to
calculate the critical temperature $T_c$ with $\sim1\%$ accuracy and
describe the critical behaviour in the 1.4\% vicinity of $T_c$ satisfying
the universality principle and the scaling relations between the critical
exponents. These features were lacking in the phenomenological theory
in \cite{beta-brass2016} based on the approximate solution of the nn
IM \cite{fisher_theory_1967}.

In the sc case it has been found that in the SC-LPA the order parameter
is accurately described within the distance $\gtrsim25\%$ away from
$T_c$. Therefore, the inability of the theory to reproduce the observed
effective critical exponent $\beta^*$ at the distance $\sim4\%$ from
the critical point has led to the conclusion that either the model
parameters are strongly influenced by the order that could exceed the
value 0.5 in this range, or that the temperatures were systematically
overestimated within the accuracy of the measurements $\sim0.2$~K or
both factors contributed to the discrepancy. Further research would be
needed to clarify this issue.

In this paper the RG equation has been derived for the simplest case
of the scalar field which is sufficient for the treatment of the Ising
model. The equation in \cite{1984}, however, was derived for the general
$n$-vector model with local interactions in the homogeneous space.
It can be easily shown that similar to the scalar case the SC-LPA RG
equation for the lattice $n$-vector models can be derived along the lines
of the present paper. In fact, this equation has already been derived in
\cite{tokar2019,tokar2019effective} within the functional renormalization
scheme in the reciprocal space of the lattice momenta. This more
complicated formalism can be necessary for the calculation of corrections
to the LPA. But within the LPA the simple approach of renormalization
in homogeneous momentum space is sufficient for recovering all concrete
results and calculations of \cite{tokar2019,tokar2019effective}.

The most serious deficiency of the LPA-based approach is that it
cannot be rigorously justified beyond the perturbation theory which is
a common problem in all strongly coupled many-body and field-theoretic
models. Strong coupling, however, is frequently encountered in physical
systems which was the reason for the development of heuristic theories
dealing with it. Arguably, among lattice models the most thoroughly
investigated are the CPA and the DMFT (see the bibliography on these
methods in review articles \cite{elliott_theory_1974,maier_quantum_2005}).
It is remarkable that, as shown in \cite{gamma_exp}, these and some
other strong coupling approximations can be derived within the same
formalism and with the effective medium self-consistency condition
similar to that used in the present paper.  This suggests that in
approximations of this kind there exists some underlying mechanism
of attenuation of the corrections.  This assumption is supported by
the excellent agreement of many experimental and MC data with the CPA
\cite{elliott_theory_1974,kissavos_total_2006,kissavos_thermodynamics_2007}
and with the SC-LPA \cite{tokar2019effective}.  Moreover, non-local
corrections to the CPA in a strongly disordered tight-binding model alloy
calculated in \cite{fiz_met_met} on the basis of the expansion suggested
in \cite{gamma_exp} were found to be in excellent agreement with the
exact MC simulations, thus justifying and improving the CPA in this
particular case. Besides, cluster generalizations of the single-site
theories have been actively developed and promising results obtained
\cite{tokar_new_1997,maier_quantum_2005,tan_topologically_2011,arXiv16}.
So there is a good deal of hope that further research along these lines
will make possible to set effective medium theories on a firm theoretical
footing.
\appendix
\section{\label{LPAapp}LPA for lattice models}
The LPA RG equation (\ref{LPA}) in the main text can be obtained from the
RG equation derived in \cite{1984} as follows. First, in the case of a
one-component field corresponding to the IM equation (8) in \cite{1984}
reads
\begin{equation}
	\frac{\partial u}{\partial\Lambda}=\frac{1}{2}
	\frac{d G}{d\Lambda}\left[\left(\frac{\Lambda}{\Lambda_{BZ}}\right)^3
		\frac{\partial^2 u}
		{\partial x^2}-\left(\frac{\partial u}{\partial x}\right)^2
	\right]
	\label{Eq8}
\end{equation}
where $u$ is the local potential, $\Lambda$ the momentum cut-off, $x$ the
local field and the propagator
\begin{equation}
	G(\Lambda) = \frac{1}{c(\Lambda)}=\frac{1}{\tilde{\epsilon}(\Lambda)+r},
	\label{G-Lambda}
\end{equation}
where $c$ is the coefficient of the quadratic in the field part
of the Hamiltonian in \cite{1984} which for easier comparison
with (\ref{G}) is convenient to separate into the dispersion
term $\tilde{\epsilon}(\Lambda)$ behaving as $\sim\Lambda^2$ when
$\Lambda\to0$ and the momentum-independent self-energy $r$. Besides,
we explicitly included in (\ref{Eq8}) the maximum cut-off momentum
$\Lambda_{BZ}$, where $BZ$ stands for the ``Brillouin zone''. In
\cite{1984} $\Lambda_{BZ}$ was set equal to unity but because in the
present paper we want to apply the equation to arbitrary lattices, the
size of BZ should also be arbitrary. Also, this factor corrects the
equation from the dimensionalities standpoint.

By substituting (\ref{G-Lambda}) into (\ref{Eq8}) one obtains the
equation that explicitly depends on the rotationally-invariant
dispersion $\tilde{\epsilon}(\Lambda)$ which according to
\cite{velicky_single-site_1968} can be fitted to the DOS of a lattice
model thus enabling application of (\ref{Eq8}) to lattice systems.
In general the fit is not unique \cite{velicky_single-site_1968} but,
fortunately, in the case of equation (\ref{Eq8}) this difficulty can be
overcome by a change of the evolution variable.  To show this let us first
divide both sides of the equation by $dG/d\Lambda$ and on the basis of
definition
\begin{equation}
	dt = \frac{dG}{d\Lambda}d\Lambda = dG
	\label{dt}
\end{equation}
introduce the new independent variable
\begin{equation}
t=G=\frac{1}{\tilde{\epsilon}(\Lambda)+r}.
	\label{t}
\end{equation}
Because all quantities here are positive, $t$ is bounded from
above by the maximum value $1/r$ reached when $\tilde{\epsilon}=0$.

In (\ref{Eq8}) $\Lambda$ is now a function of $t$ which 
formally can be found from (\ref{t}) as
\begin{equation}
	\Lambda(t) = \tilde{\epsilon}^{-1}(t^{-1}-r)
	\label{Lambda}
\end{equation}
where $\tilde{\epsilon}^{-1}$ is the function inverse to
$\tilde{\epsilon}(\Lambda)$.

The explicit dependence of $\Lambda$ in (\ref{Eq8}) on $t$
can be found with the help of the integral
\begin{equation}
	\Lambda^3(t) = \int_0^{\Lambda_{BZ}} k^3\delta[
	\tilde{\epsilon}^{-1}(t^{-1}-r)-k]dk.
	\label{Lambda3}
\end{equation}
which after integration by parts can be transformed to
\begin{eqnarray}
	\Lambda^3(t) &=& 3\int_0^{\Lambda_{BZ}} 
	k^2\theta[\tilde{\epsilon}^{-1}(t^{-1}-r)-k]
	dk\nonumber\\
	&=&3\int_0^{\Lambda_{BZ}} 
	        \theta[t^{-1}-r-\tilde{\epsilon}(k)] k^2dk,
	\label{Lambda3-2}
\end{eqnarray}
where on the first line the boundary terms were omitted by assuming
that the first term in the argument of $\theta$-function is smaller
than $\Lambda_{BZ}$ and on the second line we further assumed that
$\tilde{\epsilon}(k)$ is a monotonous function.  Though the integrand
in (\ref{Lambda3-2}) is isotropic, it can be integrated over all three
components of ${\bf k}$ by considering $\tilde{\epsilon}$ as a function of
$k=|{\bf k}|$. Now the coefficient of the second derivative in (\ref{Eq8})
can be cast in the form convenient for generalization to the lattice case:
\begin{equation}
	p(t)=\left(\frac{\Lambda(t)}{\Lambda_{BZ}}\right)^3 = 
	\frac{1}{V_{BZ}}\left.\int_{BZ}d{\bf k}
	\theta[E-\tilde{\epsilon}({\bf k})] 
	\right|_{E=t^{-1}-r}	\label{p}
\end{equation}
where $V_{BZ}=4\pi \Lambda_{BZ}^3/3$. As is easily seen, the last
expression is just the integrated DOS of the quasiparticle band with
dispersion $\tilde{\epsilon}$:
\begin{equation}
	D_{int}(E) = \int_0^{E} D(E^\prime) dE^\prime
	\label{D_int0}
\end{equation}
where $D(E)$ is the DOS corresponding to $\tilde{\epsilon}$ and, by 
construction, to $\epsilon({\bf k})$. In this way $\tilde{\epsilon}$
can be totally excluded from equation (\ref{Eq8}).

Thus, we have shown that  the rotationally invariant $G$ from \cite{1984}
and our lattice $G$ lead to the same LPA RG equation provided $D(E)$
is the same. This makes possible to establish connection between the
partition functions in both cases. By comparing our equations (\ref{Z2})
and (\ref{R}) with equation (4) in \cite{1984} for $n=1$ one sees that our
$U^b$ differs from $H_I$ in \cite{1984} only in terms that are constant in
the field and ``time'' variables. But the LPA equations depend only on the
derivatives so the constant terms in the free energy are unchanged by the
renormalization and can be accounted for at any stage.  Below they will be
gathered into one expression (\ref{u_ini}) to facilitate their analysis.

Incidentally, (\ref{p}) is also valid for $E>\max\tilde{\epsilon}$, that
is, above the upper edge of the DOS in which case the theta-function is
equal to unity so the integrated DOS of a filled band is unity.
The values of $E$ in this range are needed to integrate the
RG equation in the range where $t$ in (\ref{t}) changes from zero to
the minimum value of $G(\Lambda_{BZ}=1)$ (see Fig. 1 in \cite{1984}):
\begin{equation}
	0\leq t\leq t_0 = \min_{\Lambda} G =
(r+\max_{\Lambda}\tilde{\epsilon})^{-1} 
=[r+\max_{\bf k}\epsilon({\bf k})]^{-1}\label{0<t<G}
\end{equation} 
Because $p(t)=1$ is constant in this range, substitution 
\begin{equation}
	u=-\ln w
	\label{u-lnw}
\end{equation}
reduces the RG equation to the diffusion equation which is integrated
from $t=0$ to $t_0$ with the use of the Gaussian diffusion kernel as
\begin{equation}
	w(x,t_0)=(2\pi t_0)^{-1/2}\int dy e^{-(x-y)^2/2t_0}e^{-u^b(y)}
	\label{w}
\end{equation}
This solution is particularly useful in the IM case where according to
(\ref{R}) and (\ref{U-u}) the ``bare'' initial local potential 
\begin{equation}
	\exp[-u^{b}(x)] =\det(2\pi G)^{(1/2N)}e^{(r-\epsilon_0)/2} 
	[\delta(x-1)+\delta(x+1)]
	\label{ub}
\end{equation}
is singular and difficult to deal with numerically. Substituting
(\ref{ub}) in (\ref{w}) one gets after some rearrangement
\begin{eqnarray}
	u(x,t_0) = \frac{x^2}{2t_0}-\ln\cosh\frac{x}{t_0}-\ln2\nonumber\\
	+ \frac{1}{2}(\epsilon_0+\epsilon_{max})
	+\frac{1}{2N}\ln\det\frac{r+\epsilon}{r+\epsilon_{max}}
	\label{u_ini}
\end{eqnarray}
It is to be noted that because by assumption $\tilde{\epsilon}(k)$ and
$\epsilon({\bf k})$ have the same DOS, the maxima of both dispersions
which define its upper edge should be the same by construction. Also,
the same DOS means the same spectrum and the eigenvalues density which
means the same determinants in both cases. So in the initial condition
(\ref{u_ini}) $\tilde{\epsilon}(k)$ can be replaces by its lattice
homologue.

The usefulness of gathering all constants in $u(x,t_0)$ can be seen
from the fact that the integration range of the SC-LPA equation
$\bar{t}^R=1/r-1/(r+\epsilon_{max})$ scales as $r^{-2}$ at large $r$,
that is, in both limits $T\to\infty$ and $T\to0$. Which means that
in these limits $u(x,t_0)=u^R(x)$ so, for example, it is easy to see
using (\ref{f}) and (\ref{u_xx=0}) that in the $T\to\infty$ limit the
SC-LPA predicts the exact reduced free energy $-\ln2$. Further, by using
(\ref{f}), (\ref{h}) and (\ref{v_yy=0}) it can be shown that $m_0\to1$
when $T\to0$.  Furthermore, at large $r$ when the integration interval
is small the SC-LPA equation can be integrated as a series in $\bar{t}^R$
which can be further used to develop high- or low-temperature expansions
of the solution for comparison with known results.
\section{\label{legendre-like}The Legendre transform}
\subsection{Regularization of equation (\ref{LPA})}
To avoid dealing numerically with non-integrable singularity in the
solution (\ref{u_G}) of equation (\ref{LPA}) it was found sufficient to
slightly modify the Legendre transform for LPA equations suggested in
\cite{x-yLegendre} (see also \cite{local_potential}). The modification
consists in introducing $t$-dependence into the transform as
\begin{eqnarray}
	\label{v-u}
	v(y,t) = u(x,t) - \frac{1}{2}\bar{t}  u_x^2\\
	\label{y-x}
	y(x,t) = x - \bar{t}  u_x(x,t)
\end{eqnarray}
where $\bar{t}= t-t_0$ with $t_0$ defined in (\ref{0<t<G}). This choice
was made for convenience and in general any constant can be used instead
of $t_0$.  The independent variables in (\ref{v-u}) and (\ref{y-x})
are $x$ and $t$, $v$ and $y$ being their functions.

Now by comparing equations (\ref{v-u}) and (\ref{y-x}) differentiated with 
respect to $x$ it can be seen that
\begin{equation}
	v_y=u_x
	\label{vy-ux}
\end{equation}
if $y_x\not=0$. Similarly, by differentiating the equations 
with respect to $t$ one finds
\begin{equation}
	v_t = u_t+\frac{1}{2}u_x^2=\frac{1}{2}p(t)u_{xx}
	\label{vt-ut}
\end{equation}
where the second equality follows from (\ref{LPA}).
Finally, differentiating (\ref{vy-ux}) with respect to $x$ and substituting
$y_x$ obtained from (\ref{y-x}) one arrives at the relation
\begin{equation}
	u_{xx}=\frac{v_{yy}}{1+\bar{t} v_{yy}}
	\label{uxxvyy}
\end{equation}
which being substituted in (\ref{vt-ut}) gives the transformed RG equation 
(\ref{LPA2}) in the main text.
\section{\label{numerics}Numerical procedures}
\begin{figure}
	\centering
	\includegraphics[scale=0.8]{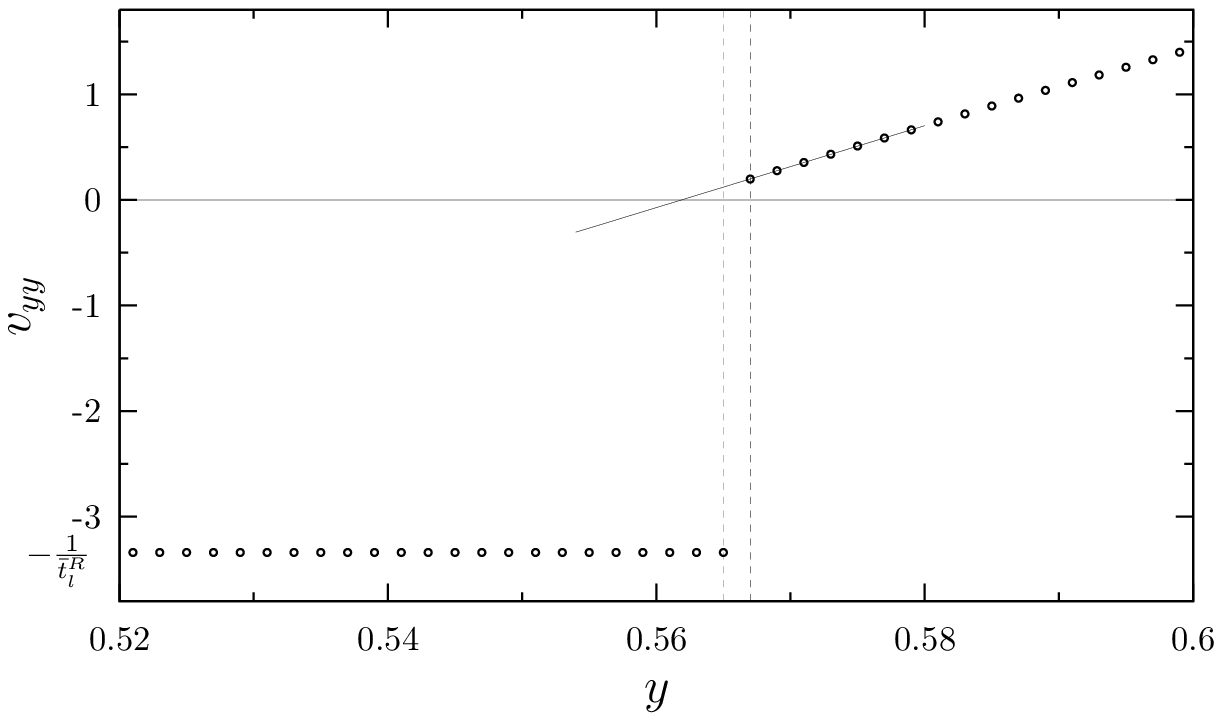}
	\caption{\label{fig7} Circles: the second
	derivative of $v(y)$ calculated for the nn sc Ising
	model below $T_c$ at an intermediate ($l$-th) iteration;
	$\bar{t}^R_l=1/r_l-1/(r_l+\epsilon_{max})$. As is seen, the
	derivative interpolated from the right of the jump interval
	bounded by vertical dashed lines does not turn to zero within
	the interval so further iterations are needed.}
\end{figure}

The evolution equation (\ref{LPA2}) has been solved by the method of lines
with the use of LSODE routine \cite{lsode} for 2500 discretization points
at the positive (due to the symmetry) $y$ axis. The point separation was
$\Delta y=2\cdot10^{-3}$ which in \cite{caillol_non-perturbative_2012}
was shown to be already small enough to give accurate values of many
quantities of interest.  In the double precision code \cite{lsode}
the use of smaller $\Delta y$ was plagued with instabilities which
restricted the accuracy of calculations of $m_0$ to $O(\Delta y)$.
The second derivatives have been approximated by the three-term central
differences in the LPA equation and by four-term one-sided differences
at the points nearest to the jump in figure \ref{fig7} with the
quadratic accuracy $O(\Delta y^2)\sim O(10^{-6})$ in both cases.  Similar
calculations performed in \cite{caillol_non-perturbative_2012} within
different renormalization schemes with the use of a quadruple precision
software showed that the accuracy can be considerably improved. Besides,
in calculations of \cite{caillol_non-perturbative_2012} the behaviour of
the second derivative of the renormalized local potential qualitatively
similar to that shown in figure \ref{fig7} was observed and its
formal and physical features discussed in detail. In the present study
we adopted the conclusion made in \cite{caillol_non-perturbative_2012}
that the discontinuity in the second derivative is physically correct
and real, though a rigorous formal proof would be desirable.

The integrated DOS needed in $p(t)$ has been calculated by numerical
integration over BZ in (\ref{p}) with $\tilde{\epsilon}$ replaced
by $\epsilon({\bf k})$. The step size in the momentum integration
was $\sim0.01$ ($\pi/300$). The integration was performed twice with
the integrand Fermi smeared at two small Fermi temperatures $T_F$ and
subsequently interpolated to $T_F=0$.  The integrations were performed
at 300 energy points and spline-interpolated in between. To improve
precision at the band edges the exactly known behaviour (\ref{k20})
was used.  The accuracy of the approximations from the renormalization
group standpoint has been checked by comparing the solutions of the LPA
equation (\ref{LPA2}) obtained with the interpolated $p(t)$ and with
the accurate analytical interpolation given in \cite{Jelitto1969609}. No
noticeable differences were found.

The solution proceeded iteratively with the self-consistent $r$ obtained
as the limit of the recursion
\begin{equation}
	r_{l+1}=r_{l}+v^R_{yy}|_{x=0} 
	\label{recursion}
\end{equation}
which converged when the self-consistency condition 
\begin{equation}
	v^R_{yy}|_{h^+=0}=0
	\label{v_yy=0}
\end{equation}was satisfied.
According to (\ref{uxxvyy}) this is equivalent to the self-consistency
condition (\ref{u_xx=0}) with $h$ in (\ref{v_yy=0}) expressed through
$y$ according to (\ref{h}).  In the symmetric phase this simply means
$y=0$ but below $T_c$ two stable solutions appear corresponding to
$y=\pm y_0\not=0$ with the spontaneous magnetisation $m_0$ given
by (\ref{m0}). So two conditions should be fulfilled below $T_c$:
(\ref{v_yy=0}) and $h=0$.
\section*{Acknowledgements}
I expresses my gratitude to Universit\'e de Strasbourg and IPCMS for
their hospitality. I am indebted to Hugues Dreyss\'e for support and
encouragement.

This research did not receive any specific grant from funding agencies in the 
public, commercial, or not-for-profit sectors.
\section*{References}
\providecommand{\newblock}{}

\end{document}